\documentclass[12pt,osajnl2,preprint,showpacs]{revtex4}  
\usepackage[draft]{hyperref} 

\begin{document}

\title{Deterministic generation of polarization-entangled photon pairs in a
cavity-QED system }

\author{Pengbo Li$^{1}$}
\author{Ying Gu$^{1}$}
\email{ygu@pku.edu.cn}
\author{ Qihuang Gong$^{1}$}
\email{qhgong@pku.edu.cn}
\author{ Guangcan Guo$^{1,2}$}
\affiliation{$^{1}$State Key Laboratory for Mesoscopic Physics,
Department of Physics, Peking University, Beijing 100871, China\\
$^{2}$Key Laboratory of Quantum Information, University of Science
and Technology of China, Hefei 230026, China}

\begin{abstract}
We propose a cavity-QED scheme that can deterministically generate
Einstein-Podosky-Rosen polarization-entangled photon pairs. A
four-level tripod atom successively couples to two high-Q optical
cavities possessing polarization degeneracy, assisted by a classical
$\pi$-polarized pump field. The stimulated Raman adiabatic passage
process in the atom-cavity system is used to produce the
polarization-entangled photon pairs. The proposal is particularly
robust against atomic spontaneous decay, which should have potential
applications in quantum information processing.
\end{abstract}

\ocis{270.5585, 270.5580.}

\maketitle

\section{Introduction}

Quantum entanglement is one of the most valuable resources in
quantum information science, which has many applications in the
fields of quantum computation and quantum communication, e.g.,
quantum cryptography, quantum teleportation and quantum
network\cite{1-Chuang}. Recently great efforts have been made to
controllably generate and detect entangled states, including the
Einstein-Podosky-Rosen (EPR) state of two qubits\cite{EPR},
Greenberger-Horne-Zeilinger (GHZ) state and W state of three qubits\cite%
{2-1,2-2}, as well as other multipartite entangled
states\cite{2-3,3,2-4}. Since photons are the ideal carriers of
quantum information, a large number of theoretical and experimental
schemes have been proposed for producing entangled photons. The
traditional optical parametric down conversion method is used to
produce the entangled photon pairs\cite{4-PRL-25-84-1970}, yet the
process is stochastic in nature. In order to controllably generate
the entangled photons, the cavity-QED scheme utilizing the coherent
interaction of atoms and field modes of a cavity is
proposed\cite{5}. Cavity QED offers an almost ideal system for the
generation of entangled states and the implementation of quantum
information processing. Experimental and theoretical progress on the
entanglement in cavity-QED with the
strong-coupling limit\cite{JPB} has been made, such as entangled atoms\cite%
{7-kimble-Xu,guo}, atom-photon entanglement\cite{8-RMP} and entangled photons%
\cite{10,JOSAB,pra2,pra3,pra4,pra5,pra6}. In this work, we have
proposed a new cavity-QED scheme composing of a four-level tripod
atom and two cavities to produce EPR entangled photon pairs. This
scheme only needs one classical pump field, therefore it may be much
easier to be implemented in experiments.

To implement the cavity-QED schemes for generating entangled
photons, one has to consider the effect of atomic spontaneous decay
on the coherent evolution of the system. This decoherence process
may be harmful  to preserving entangled state. The stimulated Raman
adiabatic passage (STIRAP) can be used to overcome this problem.
STIRAP was first used to coherently control dynamical processes in
atoms and molecules\cite{11}. It uses partially overlapping pulses
to produce complete population transfer between two quantum ground
states of an atom or molecule. In STIRAP the
population adiabatically follows the evolution of the dark state\cite%
{Luking,Li} and the excited state is never involved. Therefore, this
is particular robust against atomic spontaneous decay. The STIRAP
technique is now widely used in the chemical-reaction dynamics,
laser-induced cooling,
atom optics\cite{11} and cavity-QED systems\cite%
{prl-85-4872,kimble}. In this paper, we make use of the STIRAP
technique in a cavity-QED system composing of two cavities to
produce entangled photon pairs.

In the following, we present a cavity QED scheme which can
deterministically produce EPR polarization-entangled photon pairs. A
four-level tripod atom successively couples to two single
longitudinal mode\ high-Q optical
cavities possessing polarization degeneracy, assisted by a classical $\pi $%
-polarized pump field. The spatial profiles of the two cavity modes
and the pump field have to be overlapped, which can provide a
counterintuitive pulse sequence and maintain the two-stage STIRAP
process\cite{11}. Stage 1 is to produce a $\sigma ^{+}$ or $\sigma
^{-}$ polarized photon in cavity 1 entangled with the atom by the
first STIRAP. Stage 2 is to make the atom swap its entanglement with
the photon in cavity 1 to the photon in cavity 2 by a second STIRAP.
At this stage a two-photon polarization-entangled state is prepared
and the atom returns to its ground state. The stimulated Raman
adiabatic passage process is utilized in the cavity-QED system,
which is robust against atomic spontaneous decay. This proposal
should have potential applications in quantum information
processing.

\section{Generating entangled photon pairs by STIRAP}

The system under investigation is shown in Fig. 1. It is composed of
a four-level tripod atom and two identical high-Q cavities
possessing polarization degeneracy. The ground state of the atom is
labeled as $\left| g\right\rangle $, the two metastable states as
$\left| a\right\rangle ,\left| b\right\rangle $, and the excited
state as $\left| e\right\rangle $. The transitions $\left|
a\right\rangle \rightarrow \left| e\right\rangle $, and $\left|
b\right\rangle \rightarrow \left| e\right\rangle $ are coupled
by the cavity polarization degeneracy modes with the coupling coefficient $%
g_{i}(i=1,2)$, where $i$ denotes the $i$th cavity. The transition
$\left|
g\right\rangle \rightarrow \left| e\right\rangle $ is driven by a classical $%
\pi $-polarized pump field with Rabi frequency $\Omega $. The
detunings for
these transitions are $\Delta _{1}=\omega _{e}-\omega _{a}-\omega _{c}$, $%
\Delta _{2}=\omega _{e}-\omega _{b}-\omega _{c}$, and $\Delta
_{3}=\omega _{e}-\omega _{g}-\omega _{p}$, where $\omega _{c}$ and
$\omega _{p}$ denote the cavity mode and the pump field frequency
respectively, and $\omega _{\alpha }(\alpha =a,b,e)$ denotes the
atomic level energy. The pump field and the cavity modes have to be
overlapped spatially.{\em \ }We assume that all of them have the
Gaussian modes{\em , }i.e., $\Omega (t)=\Omega
_{0}exp[-(\frac{t-\delta t}{\Delta \tau _{p}})^{2}]$, $g_{1}(t)=g_{10}exp[-(%
\frac{t}{\Delta \tau _{c}})^{2}]$, and
$g_{2}(t)=g_{20}exp[-(\frac{t-2\delta t}{\Delta \tau _{c}})^{2}]$.
Here, $\Delta \tau _{p}$, and $\Delta \tau _{c}$ are the widths of
the pump field and the cavity mode, and $\delta t$ is the pulse
center. We focus on the situation where the two-photon resonance
happens, i.e., $\Delta _{1}=\Delta _{2}=\Delta _{3}=\Delta $. This
dark-state condition can make sure that the STIRAP in the cavity-QED
system takes place. Under the dipole and rotating wave
approximations\cite{Quantum
Optics}, the interaction Hamiltonian for this atom-cavity system is (let $%
\hbar =1$)
\begin{equation}
H_{I}=\Delta \sigma _{ee}+\Omega (t)\sigma
_{eg}+\sum_{i=1}^{2}(g_{i}(t)a_{i+}^{\dag }\sigma
_{ae}+g_{i}(t)a_{i-}^{\dag }\sigma _{be})+H.c.,
\end{equation}%
where $\sigma _{\alpha \beta }=\left| \alpha \right\rangle
\left\langle \beta \right| $, and $a_{i\pm }^{\dag }$ is the $\sigma
^{\pm }$ circularly polarized photon creation operator in the
corresponding mode. In the next two paragraphs, we give the details
of generating polarization-entangled photon pairs by STIRAP.

{\em Stage 1: Producing a photon in cavity 1 entangled with the
atom.} Suppose that the atom is initially prepared in the ground
state $\left\vert g\right\rangle $, cavity 1 in the vacuum state
$\left\vert 00\right\rangle _{1}$, and cavity 2 in $\left\vert
11\right\rangle _{2}=a_{2+}^{\dag }a_{2-}^{\dag }|00\rangle
_{2}$\cite{prl-85-4872,kimble}. There are two pathways that the atom
transfers from the ground state to the metastable
states. After undergoing the STIRAP, the atom is prepared in state $%
\left\vert a\right\rangle $ or $\left\vert b\right\rangle $ with the
same probability, and emits a $\sigma ^{+}$ or $\sigma ^{-}$
polarized photon. The interaction Hamiltonian of the cavity-atom
system at the present is
\begin{equation}
H_{I1}=\Omega (t)\sigma _{eg}+g_{1}(t)a_{1+}^{\dag }\sigma
_{ae}+g_{1}(t)a_{1-}^{\dag }\sigma _{be}+H.c.
\end{equation}%
The system has the following dark state\cite{Luking,Li}
\begin{equation}
\left\vert D_{1}\right\rangle =\{\sin \theta \lbrack \frac{1}{\sqrt{2}}%
(\left\vert a\right\rangle \left\vert 10\right\rangle
_{1}+\left\vert b\right\rangle \left\vert 01\right\rangle
_{1})]-\cos \theta \left\vert g\right\rangle \left\vert
00\right\rangle _{1}\}\otimes \left\vert 11\right\rangle _{2},
\label{dark state}
\end{equation}%
where $\tan \theta =\frac{\Omega (t)}{\tilde{g_{1}}(t)}$, $\tilde{g_{1}}(t)=%
\sqrt{2}g_{1}(t)$, $\left\vert 10\right\rangle _{1}=a_{1+}^{\dag
}\left\vert 00\right\rangle _{1}$, and $\left\vert 01\right\rangle
_{1}=a_{1-}^{\dag }\left\vert 00\right\rangle _{1}$. We then
consider the details of STIRAP process. The pulse sequence is
counterintuitive in the sense that the two initially empty levels
are coupled first, and then the initially populated level is driven
by the pump field. Moreover, the two field modes must overlap
partially. If the couplings $\Omega (t)$ and $g_{1}(t)$ change
slowly enough, and $\lim_{t\rightarrow \infty }\frac{g_{1}(t)}{\Omega (t)}=0$%
, the system will start in the state $\left\vert g\right\rangle
\otimes \left\vert 00\right\rangle _{1}\otimes \left\vert
11\right\rangle _{2}$ and end up in the state
$\frac{1}{\sqrt{2}}(\left\vert a\right\rangle \left\vert
10\right\rangle _{1}+\left\vert b\right\rangle \left\vert
01\right\rangle _{1})\otimes \left\vert 11\right\rangle _{2}$,
following the adiabatic
eigenstate given by Eq. (3). That is, when $\theta :0\rightarrow \frac{\pi }{%
2},$

\[
\left\vert D_{1}\right\rangle :\left\vert g\right\rangle \otimes
\left\vert
00\right\rangle _{1}\otimes \left\vert 11\right\rangle _{2}\rightarrow \frac{%
1}{\sqrt{2}}(\left\vert a\right\rangle \left\vert 10\right\rangle
_{1}+\left\vert b\right\rangle \left\vert 01\right\rangle
_{1})\otimes \left\vert 11\right\rangle _{2}.
\]%
As a result, the atom emits a polarized photon and is entangled with
the photon.

{\em Stage 2: The atom swapping its entanglement with the photon in
cavity 1 to the photon in cavity 2.} The atom enters cavity 2
prepared in a two-mode Fock state with just one photon in each
mode\cite{kimble}. It then interacts with the photons in cavity 2.
After undergoing the second STIRAP process, the atom absorbs one of
the photons. Now the atom swaps its entanglement with the photon in
cavity 1 to the photon left in cavity 2 and returns to the ground
state. At this stage a two-photon EPR polarization-entangled state
is prepared. The corresponding Hamiltonian described the coherent
interaction is
\begin{equation}
H_{I2}=\Omega (t)\sigma _{eg}+g_{2}(t)a_{2+}^{\dag }\sigma
_{ae}+g_{2}(t)a_{2-}^{\dag }\sigma _{be}+H.c.
\end{equation}%
In this case, the dark state is
\begin{eqnarray}
\left\vert D_{2}\right\rangle &=&\sin \beta \lbrack \frac{1}{\sqrt{2}}%
(\left\vert a\right\rangle \left\vert 10\right\rangle
_{1}+\left\vert b\right\rangle \left\vert 01\right\rangle
_{1})\otimes |11\rangle _{2}]
\nonumber \\
&&-\cos \beta \lbrack \frac{1}{\sqrt{2}}\left\vert g\right\rangle
(\left\vert 10\right\rangle _{1}\left\vert 01\right\rangle
_{2}+\left\vert 01\right\rangle _{1}\left\vert 10\right\rangle
_{2})]
\end{eqnarray}%
where $\tan \beta =\frac{\Omega (t)}{g_{2}(t)}$. If the couplings
$\Omega
(t) $ and $g_{2}(t)$ change slowly, and let $\lim_{t\rightarrow \infty }%
\frac{\Omega (t)}{g_{2}(t)}=0$, the system will begin at the state $\frac{1}{%
\sqrt{2}}\left\vert g\right\rangle \otimes (|a\rangle \left\vert
10\right\rangle _{1}+|b\rangle |01\rangle _{1})\otimes |11\rangle
_{2}$ and reach the state $\frac{1}{\sqrt{2}}\left\vert
g\right\rangle \otimes (\left\vert 10\right\rangle _{1}|01\rangle
_{2}+\left\vert 01\right\rangle _{1}\left\vert 10\right\rangle
_{2})$, following the adiabatic eigenstate by Eq. (5). That is, when
$\beta :\frac{\pi }{2}\rightarrow 0,$

\[
\left| D_{2}\right\rangle :\frac{1}{\sqrt{2}}(\left| a\right\rangle
\left| 10\right\rangle _{1}+\left| b\right\rangle \left|
01\right\rangle _{1})\otimes |11\rangle _{2}\rightarrow
\frac{1}{\sqrt{2}}\left| g\right\rangle \otimes (\left|
10\right\rangle _{1}\left| 01\right\rangle _{2}+\left|
01\right\rangle _{1}\left| 10\right\rangle _{2}).
\]%
Finally, the atom returns to its ground state $|g\rangle $, and the
two cavity photons of different polarization have been entangled
with each other. This is the central result of this work.

In order to verify the above STIRAP processes, we solve the
Schr\"{o}dinger equation numerically. The coherent dynamics of the
system is governed by
\begin{equation}
i\frac{d}{dt}|\Psi (t)\rangle =H_{I}|\Psi (t)\rangle ,
\end{equation}%
where $H_{I}$ is given in Eq. (1), and $|\Psi \rangle $ is the state
vector described the atom-cavity system. Let us consider an
alternative basis of one manifold only producing one polarized
photon in cavity 1
\begin{eqnarray}
|A\rangle &=&|g\rangle \otimes |00\rangle _{1}\otimes |11\rangle _{2}, \\
|B\rangle &=&|e\rangle \otimes |00\rangle _{1}\otimes |11\rangle _{2}, \\
|C\rangle &=&\frac{1}{\sqrt{2}}(|a\rangle |10\rangle _{1}+|b\rangle
|01\rangle _{1})\otimes |11\rangle _{2}, \\
|D\rangle &=&\frac{1}{\sqrt{2}}|e\rangle \otimes (|10\rangle
_{1}|01\rangle
_{2}+|01\rangle _{1}|10\rangle _{2}), \\
|E\rangle &=&\frac{1}{\sqrt{2}}|g\rangle \otimes (|10\rangle
_{1}|01\rangle
_{2}+|01\rangle _{1}|10\rangle _{2}), \\
|F\rangle &=&\frac{1}{\sqrt{2}}(|a\rangle |10\rangle _{1}-|b\rangle
|01\rangle _{1})\otimes |11\rangle _{2}, \\
|G\rangle &=&\frac{1}{\sqrt{2}}|e\rangle \otimes (|10\rangle
_{1}|01\rangle
_{2}-|01\rangle _{1}|10\rangle _{2}), \\
|H\rangle &=&\frac{1}{\sqrt{2}}|g\rangle \otimes (|10\rangle
_{1}|01\rangle _{2}-|01\rangle _{1}|10\rangle _{2}).
\end{eqnarray}%
It is straightforward to check that, under the two-photon resonance
condition some of the matrix elements of the Hamiltonian of Eq. (1) are $%
\langle A|H_{I}|B\rangle =\Omega (t),\langle B|H_{I}|C\rangle =\sqrt{2}%
g_{1}(t)$, $\langle C|H_{I}|D\rangle =g_{2}(t),\langle
D|H_{I}|E\rangle =\Omega (t),\langle H|H_{I}|G\rangle =\Omega
(t),\langle F|H_{I}|G\rangle
=g_{2}(t);$ while other interaction matrix elements are zero. In the basis $%
\{|A\rangle ,|B\rangle ,|C\rangle ,|D\rangle ,|E\rangle ,|F\rangle
,|G\rangle ,|H\rangle \}$, $|\Psi (t)\rangle $ has the general form
\begin{equation}
|\Psi (t)\rangle =C_{a}|A\rangle +C_{b}|B\rangle +C_{c}|C\rangle
+C_{d}|D\rangle +C_{e}|E\rangle +C_{f}|F\rangle +C_{g}|G\rangle
+C_{h}|H\rangle .
\end{equation}%
Then one can obtain the numerical solution of the system evolution.

The explicit expression for the state vector can be obtained by
solving the eigenvalue problem for the Hamiltonian. By diagonalizing
the Hamiltonian in the subspace spanned by the above five basis
states $|A\rangle ,|B\rangle ,|C\rangle ,|D\rangle $, and $|E\rangle
$, one has the dark state
\begin{equation}
|D(t)\rangle =\sin \vartheta |A\rangle -\cos \gamma \cos \vartheta
|C\rangle +\sin \gamma \cos \vartheta |E\rangle ,
\end{equation}%
where $\tan \gamma =\frac{g_{2}(t)}{\Omega (t)}$, and $\tan \vartheta =\frac{%
\tilde{g_{1}}(t)}{\sqrt{\Omega ^{2}(t)+g_{2}^{2}(t)}}$. The
two-stage STIRAP proposal can be easily seen from this dark state.
One can transfer the system from state $|A\rangle $ to $|E\rangle $
by adiabatically varying the mixing angle $\vartheta ,\gamma $. The
steps to generate the
polarization-entangled photon pair are: (i) prepare the system in state $%
|A\rangle $; (ii) change the mixing angle $\vartheta $ adiabatically from $%
\frac{\pi }{2}$ to $0$, then the system will evolve into the state
$\sin \gamma |E\rangle -\cos \gamma |C\rangle $ (stage 1), (iii)
change the mixing angle $\gamma $ form $0$ to $\frac{\pi }{2}$
slowly, the system will end up in the state $|E\rangle $ (stage 2).
It is noted that the two cavity modes and the pump field must
overlap spatially in order to maintain the adiabatic process. The
system adiabatically follows the energy eigenstate (dark state),
i.e., the system never involves the intermediate states $|B\rangle $
and $|D\rangle $, so the atomic decay is never involved. The process
including the STIRAP is very robust in producing the entangled
photons. We have to address that the two-stage STIRAP process can be
implemented by sending the atom through the two cavities. In this
case, the vacuum Rabi frequencies $g_i(t)$ can be tuned in time for
the atom to realize the STIRAP processes.

Figure 2 displays the numerical results of the Schr\"{o}dinger
equation (6). Fig. 2(a) shows the time evolution of the two cavity
coupling $g_{1}(t)$ and
$g_{2}(t)$ as well as the Rabi frequency $\Omega (t)$ of the pump field.{\em %
\ }Both the cavity modes and pump beam are assumed to have a
Gaussian
transverse shape, i.e., $\Omega (t)=\Omega _{0}exp[-(\frac{t-\delta t}{%
\Delta \tau _{p}})^{2}]$, $g_{1}(t)=g_{10}exp[-(\frac{t}{\Delta \tau _{c}}%
)^{2}]$, and $g_{2}(t)=g_{20}exp[-(\frac{t-2\delta t}{\Delta \tau
_{c}})^{2}] $. Here $\Omega _{0}=50\Gamma $, $g_{0}=10\Gamma $,
$\Delta \tau _{p}=2.5\Gamma ^{-1}$, $\Delta \tau _{c}=2.5\Gamma
^{-1}$, and $\delta
t=4.5\Gamma ^{-1}$. $\Gamma ^{-1}$ is a characteristic time, with the value $%
\Gamma \simeq 2\pi $ MHz for optical CQED and $\Gamma \simeq 2\pi $
KHz for microwave CQED\cite{8-RMP}. The above parameters come from
the recent cavity QED experiments with high finesse optical
resonators\cite{JPB,prl-85-4872} or microwave
resonators\cite{8-RMP}. In optical CQED experiments, the waists of
the cavities may be about $w_{c}\sim 20\mu m$ and the velocity of
the
atom could be $v\sim 20m/s$, then the width of cavity modes would be about $%
\Delta \tau _{c}=w_{c}/v\sim 1\mu s$\cite{JPB,prl-85-4872}. In
microwave CQED experiments, the cavity waist may be $w_{c}\sim 6mm$,
and the velocity of the atom could be $v\sim 0.5km/s$, then the
cavity modes width would be about $12\mu s$\cite{8-RMP}. The photon
life time of microwave resonators could reach $1ms$\cite{8-RMP}. The
necessary condition for adiabatic following can be maintained with
these parameters, i.e., $\Omega _{0}\Delta \tau _{p},2g_{0}\Delta
\tau _{c}\gg 1$\cite{kimble}. In Fig. 2(a), it can be seen that the
pump field and the cavity modes have been overlapped partially. The
pump field has its center displaced along the atomic beam by an
amount of $\delta t$ relative to the cavity 1 mode. The mode of
cavity 2 also has its center displaced the amount of $2\delta t$
relative to the cavity 1 mode. This pulse sequence can maintain the
adiabatic process of generating polarization-entangled photon pairs.
With the time evolution of
the cavity couplings $g_{1}(t)$, $g_{2}(t)$ and the pump Rabi frequency $%
\Omega (t)$, the whole system dynamics is shown in Fig. 2(b). The
system starts from state $|A\rangle =|g\rangle \otimes |00\rangle
_{1}\otimes |11\rangle _{2}$, via the state $|C\rangle
=\frac{1}{\sqrt{2}}(|a\rangle |10\rangle _{1}+|b\rangle |01\rangle
_{1})\otimes |11\rangle _{2}$, and eventually reaches the state
$|E\rangle =\frac{1}{\sqrt{2}}(|10\rangle _{1}|01\rangle
_{2}+|01\rangle _{1}|10\rangle _{2})\otimes |g\rangle $.
During the process, the states $|B\rangle $, $|D\rangle $, $|F\rangle $, $%
|G\rangle $, and $|H\rangle $ are never involved. In the end, the
atom returns to the ground state, and the photons existing
respectively in the two cavities have been entangled. Therefore, the
numerical simulations confirm the above STIRAP processes for
producing entangled photon pairs.

We now consider the dissipative effect on the coherent interaction
of the four-level tripod atom and two cavity modes. This includes
the spontaneous decay of atom and damping of cavity modes. As it has
been discussed previously, the STIRAP process for generating
entangled photon pairs is immune to atomic spontaneous decay. So now
only the damping of cavity modes is considered. The evolution of
density operator $\rho (t)$ in the presence
of the cavity decay is described by the master equation\cite{Quantum Optics}%
:
\begin{equation}
\label{10}
 \frac{\partial \rho }{\partial t}=-i[H_{I},\rho
]+L_{1}\rho +L_{2}\rho ,
\end{equation}%
where the cavity dissipative terms are
\begin{eqnarray}
L_{1}\rho &=&\kappa \sum_{\xi =+,-}(2a_{1\xi }\rho a_{1\xi }^{\dag
}-a_{1\xi
}^{\dag }a_{1\xi }\rho -\rho a_{1\xi }^{\dag }a_{1\xi }), \\
L_{2}\rho &=&\kappa \sum_{\xi =+,-}(2a_{2\xi }\rho a_{2\xi }^{\dag
}-a_{2\xi }^{\dag }a_{2\xi }\rho -\rho a_{2\xi }^{\dag }a_{2\xi }),
\end{eqnarray}%
$2\kappa $ is the one side decay of the two cavities, while the
other side of the cavities are assumed to be perfectly reflecting.
To solve the master equation numerically, we have used the Monte
Carlo wave function (MCWF) formalism from the quantum trajectory
methods\cite{quantum jump,cpc}. The following results are averaged
over enough realizations of quantum trajectories.

Figure 3 depicts the numerical results of the master equation (\ref
{10}) in the presence of cavity dissipation. Here the parameters are
chosen as in Fig. 2.
We consider the evolution of the system toward the entangled states $%
|E\rangle $ with the different cavity decay rate $\kappa $, i.e.,
$\kappa \simeq 0.01g_{0}$, $0.1g_{0}$ and $g_{0}$ ($g_{0}\simeq
10\Gamma $). In Fig. 3(a), the cavity decay rate is $\kappa \simeq
0.01g_{0},$ which represents
the high-Q strong coupling situation. The system starts from the state $%
|A\rangle $, evolves into the entangled state $|E\rangle $ with a
probability $p\simeq 0.80$, i.e., the success probability of
generating entangled photon pairs is $80\%$. The fidelity between
the final state and
the EPR state $F=|\langle EPR|\Psi (t=+\infty )\rangle |$ is higher than $%
90\%$, as shown in Fig. 4. In Fig. 3(b), the cavity decay rate is
$\kappa \simeq 0.1g_{0},$ which corresponds to generally strong
coupling situation. The success probability of producing the
entangled photon pairs is about $50\%$. In Fig. 3(c), the cavity
decay rate is $\kappa \simeq g_{0},$ which corresponds to the weak
coupling situation. The success probability of producing the
entangled photon pairs is neglectable.

To further gauge the performance of the scheme we plot the success
probability $P$ and fidelity $F$ as a function of $\kappa/g_0$ in
Fig. 4. When $\kappa\sim 0.1g_0-0.01g_0$, the success probability is
about $50\%-90\%$, while the fidelity is very high. The parameters
from
the recent cavity QED experiments with high finesse optical and microwave resonators are $%
(g_{0},\kappa )/2\pi \simeq (16,1.4)$ MHz \cite{nature-424},
$(g_{0},\kappa )/2\pi \simeq (16,3.8)$ MHz \cite{prl-2007}, and
$(g_{0},\kappa )/2\pi \sim (47,1)$ KHz\cite{8-RMP}, in line with the
regime of the present scheme. It can be seen from Fig. 4 that the
cavity decay strongly affects the success probability of generating
entangled photon pairs. However, the fidelity of producing the
photon pairs has only been weakly affected by cavity decay. To
generate entangled photons more efficiently, the transit time for
the atom passing through a cavity should be within the
characteristic life time of the cavity. Therefore, one has to
implement this proposed scheme in the strong coupling domain.

\section{Summary}

In summary, we have proposed a cavity quantum electrodynamics scheme
that can deterministically generate EPR polarization-entangled
photon pairs, by means of a four-level tripod atom successively
coupling to two high-Q optical cavities presenting polarization
degeneracy. This proposal relies on the cavity-QED system and
counterintuitive stimulated Raman adiabatic passage process. It is
robust against atomic spontaneous decay and should have potential
applications in quantum information processing.

\begin{acknowledgments}
This work was supported by the National Natural Science Foundation
of China under Grants Nos. 10674009, 10334010, 10521002, 10434020
and National Key Basic Research Program No.2006CB921601. The authors
acknowledge the useful discussions with professor T. C. Zhang and
Hongyan Li.
\end{acknowledgments}

\clearpage
\begin{figure}[h]
\centering
\includegraphics[bb=132 138 507 761,totalheight=6in,,clip]{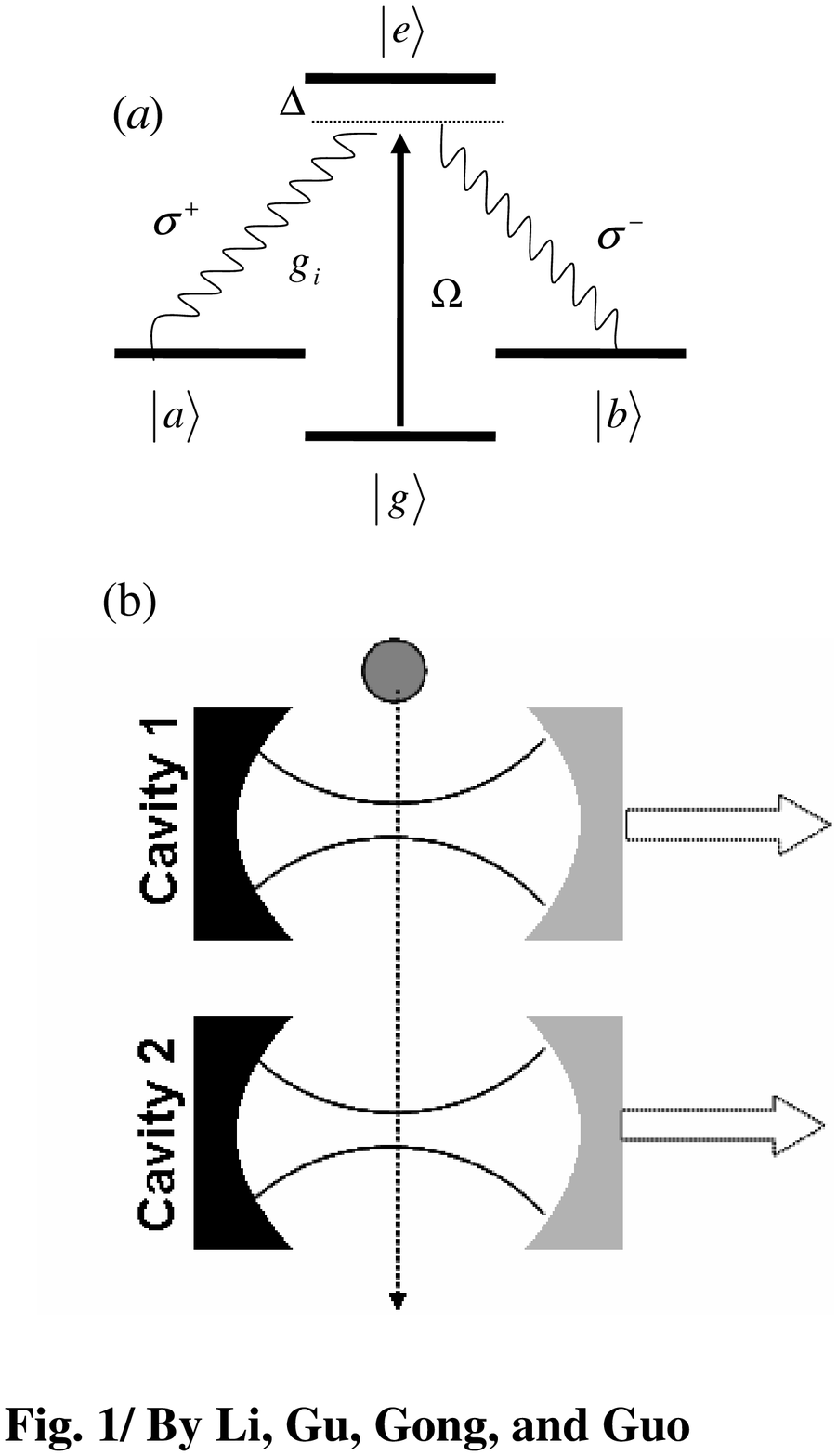}
\caption{(a) Tripod four-level atomic system under consideration.
(b) Proposed setup for the deterministic production of polarization
EPR entangled photon pairs.}
\end{figure}

\newpage
\begin{figure}[h]
\centering
\includegraphics[bb=73 153 372 713,totalheight=6in,,clip]{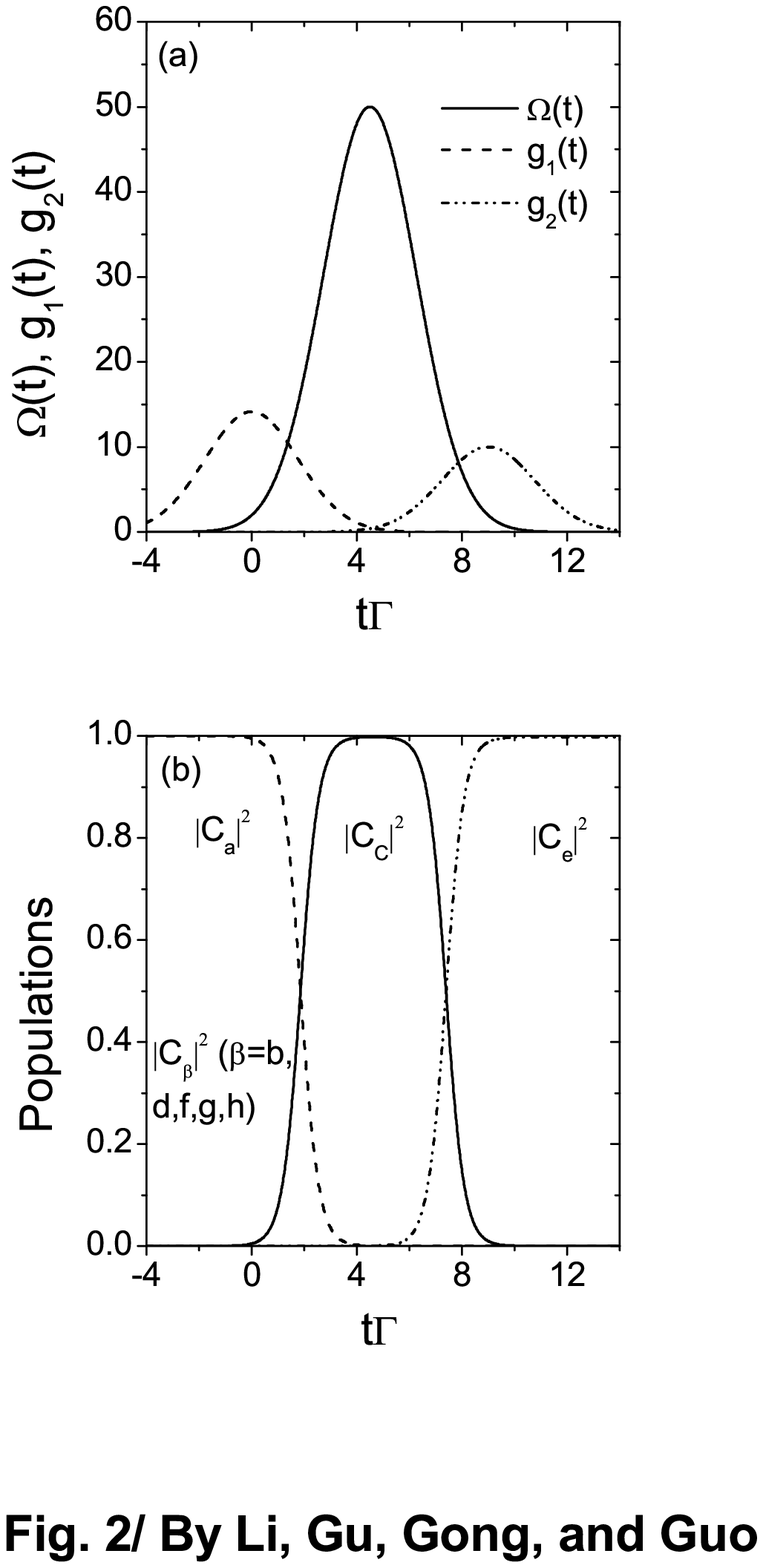}
\caption{(a) Time evolution of the coupling $g_{1}(t),g_{2}(t)$ and
Rabi
frequency $\Omega (t)$. The parameters are chosen as, $\Omega _{0}=50\Gamma$%
, $g_{0}=10\Gamma$, $\Delta\protect\tau_{p}=2.5\Gamma^{-1}$, $\Delta\protect%
\tau_{c}=2.5\Gamma^{-1}$, and $\protect\delta t=4.5\Gamma^{-1}.$ (b)
Coherent evolution of the cavity-atom system in terms of the basis
states
expansion coefficients $|C_{\protect\alpha }(t)|^{2},(\protect\alpha %
=a,b,...,h)$.}
\end{figure}

\newpage
\begin{figure}[h]
\centering
\includegraphics[bb=85 128 350 798,totalheight=7.5in,,clip]{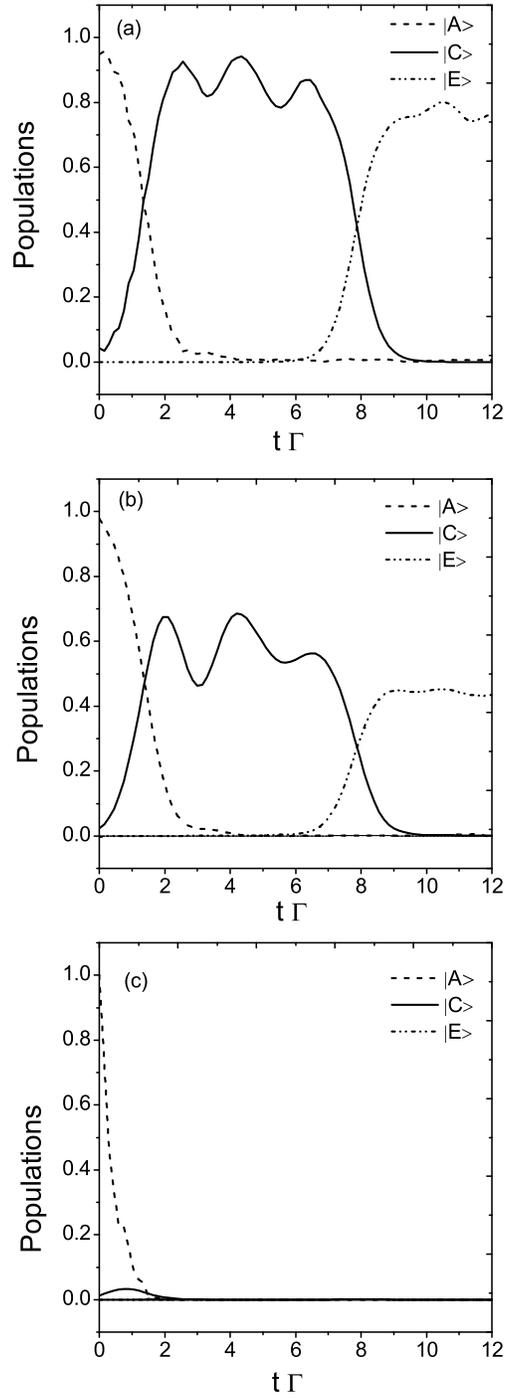}
\caption{The evolution of the system in the presence of cavity
dissipation. Parameters are chosen as in Fig. 2, but with different
cavity decay rates,
i.e., $\protect\kappa \sim0.01g_{0}$ for 3(a), $0.1g_{0}$ for 3(b), and $%
g_{0}$ for 3(c).}
\end{figure}

\begin{figure}[tbp]
\centering
\includegraphics[bb=77 182 500 575,totalheight=3.5in,,clip]{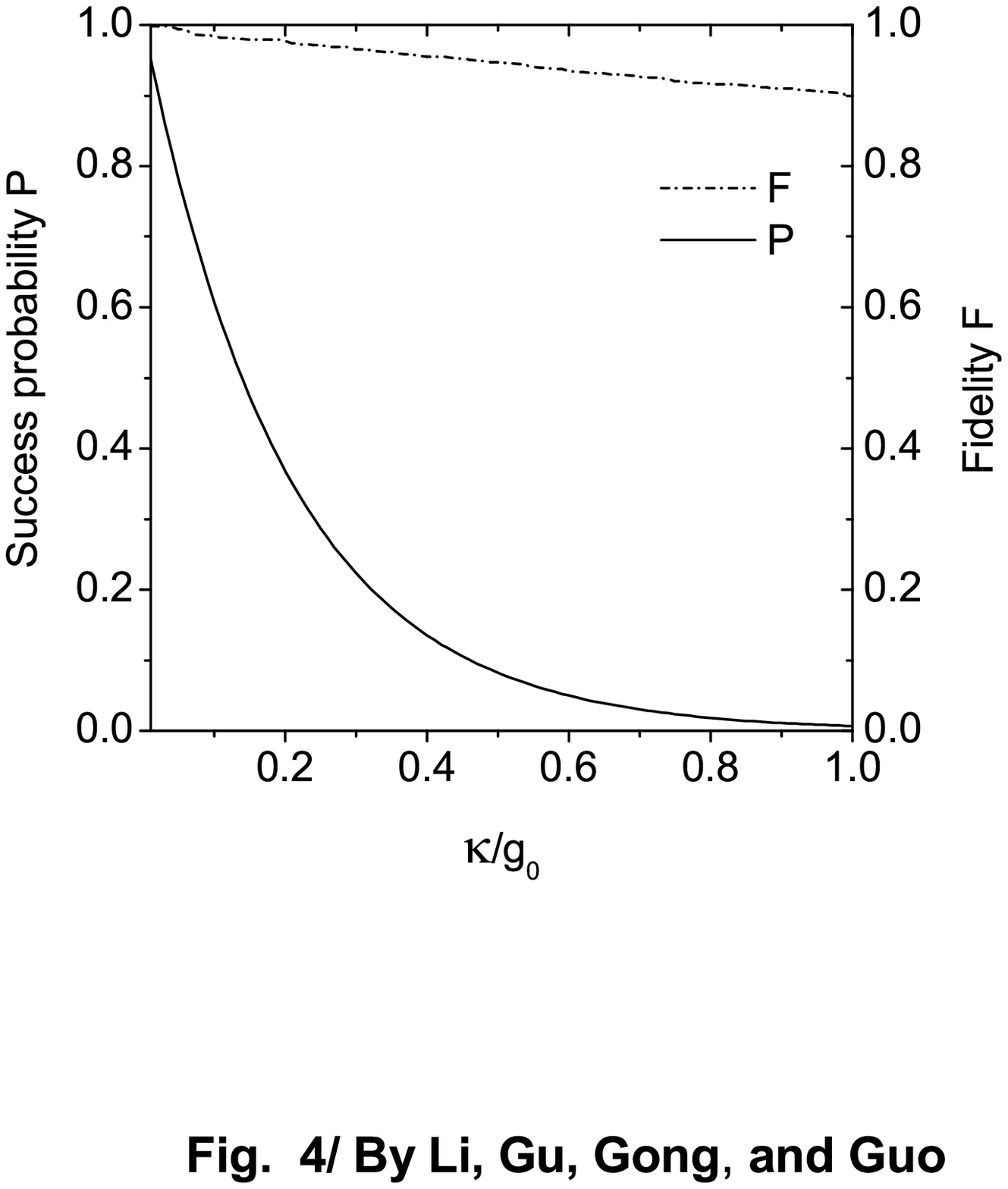}
\caption{Plots of the success probability $P$ and fidelity $F$ vs
$\kappa/g_0$, other parameters as those in Fig. 2.}
\end{figure}

\end{document}